\documentclass[a4paper,10pt,onecolumn,nofootinbib,aps,prd]{revtex4-2}
\usepackage[latin1]{inputenc}
\usepackage{amsmath}
\usepackage{amsfonts}
\usepackage{amssymb,amsthm}
\usepackage{graphicx}
\usepackage[left=2.5cm, right=2.5cm, top=3cm, bottom=3cm]{geometry}
\usepackage{float}
\usepackage{tikz}
\usetikzlibrary{shapes,arrows,shadings,shadows}
\usepackage{graphicx}
\usepackage[compat=1.1.0]{tikz-feynman} 
\usepackage{hyperref}
\usepackage[english]{babel}
\usepackage{subcaption}
\usepackage[justification=centering,singlelinecheck=false]{subcaption}
\usepackage[justification=RaggedRight,singlelinecheck=false]{caption}
\usepackage{amsfonts}
\usepackage{braket}
\usepackage{mathrsfs}
\usepackage{empheq}
\usepackage[shortlabels]{enumitem}
\usepackage{pgfplots}
\usepackage{bigints}
\usepackage{tikz, pgf}
\usetikzlibrary{decorations.pathmorphing}
\usepackage{afterpage}
\usetikzlibrary{arrows.meta}
\tikzset{%
  >={Latex[width=2mm,length=2mm]},
            base/.style = {rectangle, rounded corners, draw=black,
                           minimum width=4cm, minimum height=1cm,
                           text centered, font=\sffamily},
}
\usetikzlibrary{calc,bending,decorations.markings}
\usepackage{setspace}

\usepackage{graphicx}% Include figure files
\usepackage{dcolumn}% Align table columns on decimal point
\usepackage{bm}% bold math

\begin{document}
 	
\title{Diphoton signals for the Georgi-Machacek scenario at the Large Hadron Collider}
 	
\author{Satyaki Bhattacharya}
\email{bhattacharya.satyaki@saha.ac.in}
\affiliation{Saha Institute of Nuclear Physics \\ Sector - 1, Block- AF Bidhannagar,
Kolkata-700064 , India}
\author{Rituparna Ghosh}
\email{rg20rs072@iiserkol.ac.in}%Lines break automatically or can be forced with \\
\affiliation{Department of Physical Sciences \\ Indian Institute of Science Education and Research Kolkata \\ Mohanpur, Nadia - 741246, India}%
\author{Biswarup Mukhopadhyaya}%
\email{biswarup@iiserkol.ac.in}
\affiliation{Department of Physical Sciences \\ Indian Institute of Science Education and Research Kolkata \\ Mohanpur, Nadia - 741246, India}%
 	
\begin{abstract}The diphoton channel for exploring the Georgi-Machacek (GM) scenario containing scalar triplets at the Large Hadron Collider (LHC) has been identified as germane and subjected to a detailed 
study. The scalar spectrum of the model, which imposes a custodial SU(2) on the potential, gets classified into a 5-plet, a 3-plet
and two singlets under the custodial symmetry. While most attempts to probe or constrain the scenario at the LHC depend largely on signals of charged scalars, we point out that the
custodial SU(2) singlet state $H$ can have a substantial branching ratio (amounting to a few percent) into two photons. We carry out a detailed simulation of the resulting signal and the standard model backgrounds, obtaining the signal significance in different regions of the parameter space using the profile likelihood ratio method.  Substantial regions of the GM parameter space are thus shown to be accessible to LHC studies, both at the high-luminosity
run with $\int{\cal L} dt = 3000 fb^{-1}$, and also in Run-3 with  $\int{\cal L} dt = 300 fb^{-1}$,
even after folding in systematic errors. We have also demonstrated that a rather substantial improvement in the signal significance is achieved by switching over from a cut-based analysis to one based on a neural network.
\end{abstract}
\maketitle

%\large
\newpage
\section{Introduction}
\label{intro}
Scenarios where electroweak symmetry is broken by the vacuum expectation values (vev) of scalars
extending beyond the single SU(2) doublet Higgs of the Glashow-Salam-Weinberg theory \cite{Glashow}-\cite{Weinberg}, are not only consistent but also of considerable phenomenological interest. Among them, two Higgs doublet models (2HDM) \cite{2hdm1} - \cite{Branco} are the most widely studied ones. It is, however, also possible for scalars belonging to higher representations of $SU(2)$ to have a role in electroweak symmetry breaking (EWSB). The existence of scalar triplets is such a possibility; they can play a role
in generating $\Delta L = 2$ mass terms for neutrinos, leading to the Type-II seesaw mechanism \cite{Gu}-\cite{Primulando}.
The triplet vev, however, is restricted by the $\rho$-parameter (defined  as $\frac{m_w^2}{m_z^2 \cos^2\theta_w}$) to be $\leq 4.8$~ GeV \cite{Primulando}\cite{Ghosh}.
In such a case, the role of the triplet(s) in EWSB is hardly palpable. The question one can ask in
this context is: can scalar triplets occur in such a way that the triplet vev can be
big enough  to have a role in accelerator phenomenology involving the electroweak sector? 
\vspace{0.1cm}

This is indeed possible in a class of Higgs triplet models, of which the Georgi-Machacek
(GM) scenario is by far the best-known one \cite{GM}. Here, a complex (Y = 2) scalar triplet $\chi$ and a real (Y = 0) scalar triplet $\xi$ 
are postulated, over and above $\Phi$, the complex (Y = 1) doublet of the standard model (SM). A custodial
global $SU(2)$ symmetry is imposed on the scalar potential, which ensures $\rho = 1$ at the tree-level, with $v_\chi = v_\xi$, where $v_\chi$ and $v_\xi$ are the complex and real triplet vev respectively. The custodial symmetry is robust against radiative corrections involving scalars, although corrections involving $U(1)_Y$ gauge couplings can break it \cite{CG} \cite{wud2}. The resulting fine-tuning required for maintaining the 
$\rho$-parameter within its constraints, is not worse than that involved in stabilizing the Higgs mass in the SM itself. 
Keeping this in mind, several studies have been carried out on the
collider signals and other implications of such a scenario, some of which constrain the parameters of the model \cite{bmz} - \cite{Ghosh2}. One important finding emerging from such studies is that, in general, triplet vevs up to about 40 GeV are consistent with all current constraints, and the phenomenology of the scenario, including constraints on it from accelerator data, does not change appreciably if the 
extent of custodial symmetry breaking leads up to $\simeq$ 30\% splitting between
the real and complex triplet vevs \cite{Ghosh2}. In the present study, we point out the usefulness
of the diphoton signal at the Large Hadron Collider (LHC), hitherto unexplored in the context of this scenario, in probing regions of its parameter space not constrained so far. We also show that one may be standing at the doorstep of probing the GM scenario via the heavier scalar $H$ decaying to a pair of photons, if an analysis based on neural networks is carried out.
\vspace{0.1cm}

Under the custodial SU(2), the physical scalar states in the GM scenario get classified into a 5-plet, a 3-plet, and two electrically neutral
singlets. One of the singlets is the SM-like state $h$, while the other state is denoted by $H$. $h$ is already well-studied at the LHC, mostly
through final states arising from its decay into fermion and gauge boson pairs. The decay channel $h\rightarrow\gamma\gamma$ provides useful information,
although the branching ratio is $\simeq 10^{-3}$\cite{ATLAS15diph}-\cite{CMS21diph}. For $H$, on the other hand, the suppression of the coupling strength with fermion pairs as well as with $WW, ZZ$ causes enhancement of the $\gamma\gamma$ branching ratio, raising it to the level of a few percent to about 10\%. This happens in regions of the parameter space
which are otherwise allowed by current analyses, based mainly on the production and decays of the doubly charged scalar $H_5^{\pm\pm}$ of the GM
model. It is thus important to investigate if the diphoton decay channel of $H$ can extend the regions that can be probed at the LHC, especially in its high-luminosity run.
Our study reveals an answer in the affirmative direction, especially in regions
where the mass of $H$ is in the range of 160-250 GeV.
\vspace{0.1cm}

Apart from some indirect constraints coming from the charged scalar belonging to the 3-plet of the custodial SU(2), which contributes to rare decays like $b\rightarrow s\gamma$ and $B_s \rightarrow \mu^+ \mu^-$ \cite{Indir}, the bulk of current phenomenological
constraints on the GM model is based on the analysis of data available from an integrated luminosity of about $139 fb^{-1}$ 
at the LHC \cite{ATLAS_DY},\cite{atlsdiph}. The 5-plet plays a crucial role here, primarily through production and decay of the doubly charged
scalar $H_5^{\pm\pm}$. Earlier studies set such constraints based on the assumption that the decay $H_5^{++} \rightarrow W^+W^+$
dominates \cite{logan_const2} \cite{ATLAS-doub} \footnote {In principle, the decay into same-sign dileptons can lead to powerful probes. But that channel has a significant role for triplet vev $\leq 10^{-6}$ GeV in order to be consistent with neutrino mass generation via type-II seesaw mechanism. Therefore, this channel is not of much use in probing regions of the parameter space where the vev is around the GeV scale, where the triplets play significant roles in collider phenomenology.}. It was shown more recently that it is possible to have more relaxed upper limits on the triplet vev
if one takes into account the additional decays $H_5^{++} \rightarrow H_3^+W^+, H_3^+ H_3^+$ \cite{Ghosh2}. The custodial 3-plet $(H_3^\pm, H_3^0)$, on the other
hand, leads to signals that have sizeable backgrounds, and the probability of being faked by signals of 2HDM is also high. It is therefore a notable observation that the heavier neutral scalar, singlet under the custodial SU(2), can be the source of substantial diphoton signals. The allowability of relatively low-mass charged scalars in the GM scenario, which causes enhancement of loop amplitudes, is one feature that is responsible for the enhancement of diphoton signals in the corresponding regions of the parameter space. \newline
\vspace{0.1cm}

In short, the novelty of the present work lies in the following observations:\newline
1. The GM parameter space admits of hitherto unexplored regions where the heavier neutral scalar $H$ decays into diphoton final states with substantial statistical significance.\newline
2. The detectability of the signal is demonstrated through both a cut-based analysis and one based on a neural network. The reported signal significance has been computed using the binned profile likelihood ratio method.\newline
3. The diphoton signals are of importance, not only in a multichannel analysis aimed at unraveling a GM-like scenario, but also in those cases when the signal coming from $H^{\pm\pm} \rightarrow W^{\pm}W^{\pm}$ is diluted. This can happen for (a) small triplet vev, and (b) cases where the branching ratio for $H^{\pm\pm} \rightarrow W^{\pm}W^{\pm}$ gets diluted by decays into $W^{\pm}H_3^{\pm}$ and $H_3^{\pm}H_3^{\pm}$.

The paper is organised as follows: in Sec. \ref{model}, we present a brief outline of the model and point out the reason for the occurrence of diphoton signals from the neutral scalar $H$ in the mass range 160-250 GeV. Sec. \ref{bp} focuses on different constraints on the parameter space coming from both theoretical and experimental considerations. The methodology we have followed to fix our benchmark points has also been discussed there. Sec. \ref{sigbkg} is devoted to a detailed discussion of the simulation of background and signal events. Finally, in Sec. \ref{rslt}, we give an outline of the profile likelihood ratio method, we used to calculate the signal significance and then proceed to present our final results in terms of parameter regions of the model. We summarize and conclude in Sec. \ref{cncl}. 
%\vspace{-1cm}

\section{A brief outline of the model}
\label{model}
\vspace{0.2cm}

In the GM model, the scalar sector of the SM has been extended with a $Y=2$ complex triplet $\chi = (\chi^{++},\chi^{+},\chi^{0})^{T}$ and a $Y=0$ real triplet $\xi = (\xi^{+},\xi^{0},\xi^{-})^{T}$. To manifest the custodial symmetry in the scalar potential, the complex triplet $\chi$ and the real triplet $\xi$ are combined as a $SU(2)_{L} \times SU(2)_{R}$ bi-triplet $X$ which transforms under the aforementioned group as $X \rightarrow U_L X U_R^\dagger$. The Standard model Higgs doublet is also presented as a bi-doublet $\Phi$.
\begin{equation}
		\Phi = \begin{pmatrix}
		\phi^{0 \star} & \phi^+\\
		\phi^- & \phi^0 
		\end{pmatrix} \ \ \
		X = \begin{pmatrix}
						\chi^{0 \star} & \xi^+ & \chi^{++}\\
						\chi^- & \xi^0 & \chi^+\\
						\chi^{--} & \xi^- & \chi^0\\
			\end{pmatrix} \ \ \
	\end{equation}
The scalar potential of the model as a function of $\Phi$ and $X$ is given by,
\begin{eqnarray}
\label{pot}
					V(\Phi,X)&=&\frac{\mu_2^2}{2} Tr(\Phi^\dagger\Phi) + \frac{\mu_3^2}{2}Tr(X^\dagger X) + \  \lambda_1[Tr(\Phi^\dagger\Phi)]^2 \nonumber \\
                    &+& \  \lambda_2Tr(\Phi^\dagger\Phi)Tr(X^\dagger X)
					+ \  \lambda_3Tr(X^\dagger X X^\dagger X) \nonumber \\
                    &+& \  \lambda_4[Tr(X^\dagger X)]^2 - \  \lambda_5Tr(\Phi^\dagger \tau^a \Phi \tau^b)Tr(X^\dagger t^a X t^b) \nonumber \\
                    &-& M_1Tr(\Phi^\dagger \tau^a \Phi \tau^b)(UXU^\dagger)_{ab} - M_2Tr(X^\dagger t^a X t^b)(UXU^\dagger)_{ab}
\end{eqnarray}
where $\tau^a = \frac{\sigma^a}{2}$, $\sigma^a, (a=1,2,3)$ being the Pauli matrices and $t^a, (a=1,2,3)$ are the $SU(2)$ generators in the triplet representation. The matrix $U$ is defined as,
\begin{equation}
U = \begin{pmatrix}
    -\frac{1}{\sqrt2} & 0 & \frac{1}{\sqrt2}\\
    -\frac{i}{\sqrt2} & 0 & -\frac{i}{\sqrt2}\\
    0 & 1 & 0\\
    \end{pmatrix}
\end{equation}

This specific form of the potential ensures that the minimization conditions always have a solution where the complex triplet vev and the real triplet vev are equal. With the triplet and doublet vev denoted by $v_\chi$ and $v_\phi$ respectively, the potential minimization conditions are given by,
\begin{eqnarray}
\label{min1}
&\mu_2^2 v_\phi& + \ 4  \lambda_1 v_\phi^3 + 3 (2  \lambda_2 -  \lambda_5 ) v_\phi v_\chi^2 - \frac{3}{2}M_1 v_\phi v_\chi = 0 \\
\label{min2}
&3\mu^2_3& v_\chi + \ 3 (2  \lambda_2 -   \lambda_5 ) v_\phi^2 v_\chi + 12 (  \lambda_3 + 3  \lambda_4 ) v_\chi^3 - \frac{3}{4}M_1 v_\phi^2 - 18M_2 v_\chi^2 = 0
\end{eqnarray}

Since the vacuum retains a custodial $SU(2)$ after EWSB, the mass eigenstates at the tree-level form multiplets under the custodial symmetry. The Goldstone modes $(G^\pm, G^0)$, which constitute a triplet under the custodial $SU(2)$, are eaten up by the weak bosons $W$ and $Z$ and the physical particle spectrum consists of a 5-plet $({H_5^{\pm\pm},H_5^{\pm},H_5^{0}})$, a 3-plet $({H_3^{+},H_3^{0},H_3^{-}})$ and two CP-even singlets ${H,h}$. Other than scalar self-interactions, the 5-plet couples to gauge boson pairs only, while 3-plet interacts only with fermion pairs. We have fixed $h$ to be the 125 GeV scalar with $H$ being the heavier singlet state and the source of the excess diphoton, we are studying here.

\subsection{\large{An additional source of diphoton events}}

  Since this work focuses on diphoton excess coming from the scalar sector of the GM model, here we will demonstrate the source diphoton in this context. For this let us review the composition of the states. The 5-plet states are composed of triplets only,
\begin{equation}
    H_5^{++} = \chi^{++} \ , \ \ 
    H_5^+ = \frac{\chi^+ - \xi^+}{\sqrt{2}} \ , \ \ 
    H_5^0 = \sqrt{\frac{2}{3}}\xi^0 - \sqrt{\frac{1}{3}}\chi^{0,r}
\end{equation}
 
The 3-plet states, carrying a pseudoscalar $H_3^0$, are given by,
\begin{equation}
    H_3^{+} = -s_H\phi^+ + c_H\frac{\chi^+ +  \xi^+}{\sqrt{2}} \ \ , \ H_3^0 = -s_H\phi^{0,i} + c_H\chi^{0,i}
\end{equation}
where,
\begin{equation}
  s_H = \frac{2\sqrt{2}v_\chi}{v} \ ,\ \ c_H = \frac{v_\phi}{v}    
\end{equation}
 
Lastly, the two CP-even singlets are,
\begin{equation}
    H^0 = \phi^{0,r} \ \ , \ H^{\prime 0} = \sqrt{\frac{1}{3}}\xi^0 + \sqrt{\frac{2}{3}}\chi^{0,r}
\end{equation}

The physical neutral scalar states are in general linear superposition of the above two CP-even singlets and are given by,
\begin{equation}
    h = \cos{\alpha}\phi^{0,r} - \sin{\alpha}H^{\prime 0} \ , \ \ H = \sin{\alpha}\phi^{0,r} + \cos{\alpha}H^{\prime 0}
\end{equation}
The angle $\alpha$ depends on the $2 \times 2$ CP-even custodial-singlet scalar mass matrix. The elements of the mass matrix are,
\begin{eqnarray}
\label{m11}
\mathcal{M}_{11}^2 &=& 8\  \lambda_1 v_\phi^2 \\
\mathcal{M}_{12}^2 &=& \frac{\sqrt{3}}{2} v_\phi[-M_1 + 4(2\lambda_2 - \  \lambda_5)v_\chi] \\
\label{m22}
\mathcal{M}_{22}^2 &=& \frac{M_1 v_\phi^2}{4 v_\chi} - 6M_2v_\chi + 8(\lambda_3 + 3\  \lambda_4)v_\chi^2 \\
\tan2\alpha &=& \frac{2\mathcal{M}_{12}^2}{\mathcal{M}_{22}^2 - \mathcal{M}_{11}^2}
\label{matrix}
\end{eqnarray}

In addition to $h$, $H_5^0$, $H_3^0$, and $H$ can decay into diphoton. As $H_5^0$ couples to gauge bosons only, at the LHC it will be produced mostly via the Vector Boson Fusion(VBF) channel. Hence, for the mass range $m_5 \geq 160 $ GeV, the $W$ boson decay mode will be the dominant decay channel, and there will be no scope to enhance the diphoton decay mode, keeping the production cross-section fixed. Also, the charged scalar loop will not contribute much due to $H_5^{++}$ belonging to the same multiplet as $H_5^0$. Hence the diphoton decay of $H_5^0$ will not bear any fruitful result in the collider context in the parameter regions where triplet vev is substantial. On the other hand, top loop is the only contributor for $H_3^0$ decaying into diphoton, and hence here also no enhancement in BR($H_3^0 \rightarrow \gamma \gamma) $ is possible.  
\vspace{0.1cm}

For the state $H$, since it couples to both fermions and gauge bosons, it can be produced via gluon-fusion (ggF) mode which has a higher cross-section than VBF at the collider. In this context, the most crucial part is the branching ratio of $H \rightarrow \gamma \gamma$. In addition to the fermion and gauge boson loops, here charged scalar loops contribute significantly to this decay mode, enhancing the branching ratio up to $\approx 10\%$ in the mass range $160-250$ GeV. Hence, in spite of suppression on the production side as compared to $h$-production
in the gluon fusion channel, this $H \rightarrow \gamma \gamma$ decay mode provides a hopeful path  to be probed at the collider at least at the high luminosity run.\footnote{The branching ratio sometimes tends to be close to 10\%, mainly for the following reasons.\newline
	(a) For larger $M_2$, the triplet component in $H$ pushes up the triliear couplings in the loop.\newline
	(b) The $WW$ and $ZZ$ decays of an $H$ are suppressed for low $s_H$. On the whole, however, the enhancement of the diphoton signal depends also on $\sigma(p p \rightarrow H )_{ggF}$, which is often anti-correlated with BR$(H\rightarrow \gamma \gamma)$. These features are all reflected in the benchmarks listed in Table~\ref{sgbp}}.
\vspace{0.1cm}

Such diphoton signals  have been  mentioned in some recent work \cite{Bairi}, but without full simulation and detailed assessment of the backgrounds. People have also invoked the GM scenario as a source of diphoton, but in the exclusive context of a claimed excess around 95 GeV \cite{95gev1} \cite{95gev2}. Our analysis, on the other hand, pertains to excess diphoton across the GM parameter space, especially in the slightly higher mass regions, where the backgrounds are less susceptible to uncertainties and the potential of this scenario can be explored more widely.
\vspace{0.1cm}

We end this section with a brief overview of the couplings contributing to this search. As already emphasized, scalar loops play a pivotal role in enhancing the $H \rightarrow \gamma \gamma$ branching ratio. Due to the custodial $SU(2)$ symmetry of the potential, both the $HH^{\pm}_5H^{\mp}_5$ and $HH^{\pm\pm}_5H^{\mp\mp}_5$ coupling strengths are the same, and we will denote it as $g_{HH_5H_5}$. Similarly, $g_{HH_3H_3}$ represents the coupling strength of $H$ to an $H^{\pm}_3$-pair. 
In terms of parameters in the scalar potential,
\begin{eqnarray}
\label{h5c}
    g_{HH_5H_5} &=& 8 \sqrt{3}(\lambda_3+\  \lambda_4)v_{\chi}\cos{\alpha} + (4\lambda_2 + \  \lambda_5)v_{\phi}\sin{\alpha} + 2\sqrt{3} M_2\cos{\alpha}
\end{eqnarray}
\begin{eqnarray}
\label{h3c}
     g_{HH_3H_3} &=& 64 \lambda_1  \frac{v_{\chi}^2                     v_{\phi}}{v^2} \sin{\alpha} + \frac{8}{\sqrt{3}} (\lambda_3 + 3 \lambda_4) \frac{v_{\phi}^2 v_{\chi}}{v^2} \cos{\alpha}   \nonumber \\
                 &+& \frac{4}{\sqrt{3}} M_1 \frac{v_{\chi}}{v^2} (v_{\chi} \cos{\alpha} + \sqrt{3}             v_{\phi} \sin{\alpha} ) \nonumber \\
                 &+& \frac{16}{\sqrt{3}}(6 \lambda_2 + \lambda_5)\frac{v_{\chi}^3}{v^2} \cos{\alpha}
                      + (4   \lambda_2 - \lambda_5) \frac{v_{\phi}^3}{v^2} \sin{\alpha} \nonumber \\
                 &-& 2\sqrt{3}M_2 \frac{v_{\phi}^2}{v^2} \cos{\alpha}
                    + \frac{8}{\sqrt{3}} \lambda_5 \frac{v_{\chi}v_{\phi}}{v^2}
                    (v_{\phi} \cos{\alpha} + \sqrt{3}      v_{\chi} \sin{\alpha})
\end{eqnarray}
The Yukawa coupling of $H$ is given by,
\begin{eqnarray}
    y_{H\bar{f}f} = m_f\frac{\sin{\alpha}}{v_\phi}
\end{eqnarray}
with $m_f$ being the fermion mass.

Since we are looking at the diphoton channel, the doubly-charged scalar loop has the greatest influence. As there are strong bounds on quartic couplings from theoretical considerations like unitarity and vacuum stability, they cannot contribute much to increase the diphoton branching ratio of $H$. Here the trilinear couplings play a significant role in increasing the contribution of charged scalar loops, giving rise to a sizeable diphoton signal at the LHC. Hence, there is no such signal present for the case of the $Z_2$ symmetric GM model \cite{wud1} \cite{loganz2}. As $H$ is the triplet-dominated state, and the doubly charged scalar is also composed of complex triplet only, higher values of the trilinear coupling $M_2$ drive better signal strength. Also, the signal is prominent mostly in the regions with $m_5 \leq m_3$, where $m_5$ is the mass of the 5-plet state and $m_3$ is the mass of the 3-plet state. Thus this decay channel not only opens a new path to be explored at the LHC but also differentiates the aforementioned two types of the GM scenario and also hints at a specific mass hierarchy within the scalar sector of the model. 

\vspace{-1cm}

\section{Benchmarks and constraints}
\label{bp}
The benchmark points in the GM parameter space pertaining to our study have been selected to highlight regions when the diphoton signal strength is expected to be most prominent. Such regions, of course, have to be consistent with all theoretical and phenomenological constraints. As would be demonstrated in the next section, the benchmarks worth studying turn out to correspond to $m_H$ in the range 160-250 GeV.\newline
The diphoton event rate via $H$, without taking into account the effects of cuts is given by,
\begin{equation}
    \sigma(pp \rightarrow H)_{ggF} \times BR(H \rightarrow \gamma \gamma) = [\zeta_{F}^{2} BR(H \rightarrow \gamma \gamma)] \ [\sigma(pp \rightarrow H)^{SM}_{ggF}]_{m_h^{SM} = m_H}
\end{equation}
where $\zeta_F$ represents the coupling strength modification factor for the scalar state $H$ to charged fermion-pairs with respect to the SM Yukawa coupling.
\begin{eqnarray}
	\label{zetaf}
    \zeta_F &=& \frac{y_{H\bar{f}f}}{y_{h^{SM}\bar{f}f}} \ = \frac{\sin \alpha}{\sqrt{1 - \frac{8 v_\chi^2}{v^2}}}
\end{eqnarray}
$h^{SM}$ being the SM Higgs boson and $y_{h^{SM}\bar{f}f}$ being its Yukawa coupling. The quantity $\zeta_F^2 BR(H \rightarrow \gamma \gamma)$ encapsulates the required model information for the signal process.
Following equation \eqref{zetaf}, $\zeta_F^2$  represents the modification of the $H$-production cross-section via gluon-gluon fusion with respect to the similar cross-section for an SM-like Higgs with mass $m_H$.
\vspace{0.1cm}

The input set to scan the parameter space consists of ($m_H$,$m_5$,$m_h$,$s_H$,$\lambda_1$,$\lambda_3$,$\lambda_4$,$M_2$), where we have fixed $m_h$ at 125 GeV. In terms of these parameters, the other potential parameters are given by,
\begin{eqnarray}
    M_1 &=& \frac{4v_\chi}{v_\phi^2}[m_H^2 + m_h^2 - 8\lambda_1 v_\phi^2 - 8(\lambda_3 + 3\lambda_4)v_\chi^2 + 6M_2v_\chi] \\
    \lambda_5 &=& \frac{2}{3v_\phi^2}(m_5^2 - m_H^2 - m_h^2 + 8\lambda_1v_\phi^2 + 24\lambda_4v_\chi^2 - 18M_2v_\chi) \\
    M_{12}^2 &=& \sqrt{M_{11}^2 M_{22}^2 - m_H^2 m_h^2}  \\
    \lambda_2 &=& \frac{1}{2}[\lambda_5 + \frac{1}{4v_\chi}(M_1 - \frac{2M_{12}^2}{\sqrt{3}v_\phi})] 
\end{eqnarray}
Where, $M_{11}^2$ and $M_{22}^2$ are given by equations \ref{m11} and \ref{m22} respectively. For a fixed $m_H$, we scan over different values for each of the remaining six parameters and the point where $\zeta_{F}^{2} BR(H \rightarrow \gamma \gamma)$ attains a maximum, marks our benchmark point (BP) for the signal. The ranges over which we scan the six parameters are,
\begin{align}
&m_5 \ \in (m_H - 60) \ \rm GeV  \ to \ (m_H -40) \ \rm GeV \nonumber \\ 
&s_H \in (0.1 - 0.4) \nonumber \\
&\lambda_1 \in (0.03 - 0.06) \nonumber \\ 
&\lambda_3, \lambda_4 \in (-1.5,1.5) \nonumber \\ 
&M_2 \in (20 - 100) \ \rm GeV
\end{align}
A lower limit on $s_H$ in the scan is kept because our emphasis is on those situations where the triplet vev has a significant role in EWSB. $s_H \ge 0.4$ is disfavoured by searches for $H_5^{\pm\pm}$ in $W^{\pm}W^{\pm}$ decay channel.\newline
$m_5 < m_H -60$ is disfavoured by the custodial $SU(2)$. On the other hand, $m_5 > m_H-40$ leads to suppression of the loop amplitudes, unless $M_2$ is large and $s_H$ is low, in which case again $H$ will have an enhanced triplet component, thus reducing the Yukawa-driven production rate.\newline
$M_2 > 100$ GeV is unlikely to enhance the diphoton signal, due to the reason just stated above. $M_2 \le 20$ GeV, on the other hand, suppresses the trilinear scalar coupling so much that the diphoton branching ratio suffers.\newline 
The BPs are required to satisfy three types of constraints on the parameter space: theoretical constraints, indirect constraints, and constraints coming from the measurements of the 125 GeV scalar and searches for  beyond standard model (BSM) particles in the EWSB sector.

\vspace{-1.2cm}

\subsection{\large{Theoretical constraints}}
The theoretical constraints come from perturbative unitarity and the demand for a global stable EWSB vacuum respecting custodial $SU(2)$. For details, the reader is referred to the relevant literature \cite{DECOUP}.

\subsection{\large{Indirect constraints}}
Indirect constraints come from electroweak precision tests i.e. precision measurements of $S$, $T$, and $U$ parameters \cite{Peskin1} \cite{Peskin2} and measurements of some rare decay processes such as $b \rightarrow s \gamma$, $B_s \rightarrow \mu^+ \mu^-$. Since the charged scalar $H_3^+$ contributes to the rates of these rare decays, this in turn puts a constraint on $m_3$ and $c_H$.\newline
Both theoretical and indirect constraints on the parameter space are applied via GMCALC-1.5.3 \cite{Indir},\cite{DECOUP},\cite{loopdecay}.

\subsection{\large Constraints from direct searches of BSM scalars and measurements of 125 GeV scalar}
It has been ensured that each BP is consistent with all the LHC data available till now i.e. latest results from the Run-2 data with $\int\mathcal{L}dt = 139 \ fb^{-1}$. We have imposed the constraints from direct searches for a singly charged Higgs and any non-standard neutral Higgs, using the HiggsBounds \cite{hb1}-\cite{hb5} module of HiggsTools\cite{htool}. The constraints coming from measurements on the $125$ GeV scalar, is applied via the module HiggsSignals \cite{hs1} \cite{hs2} built within HiggsTools. %The points are chosen such that $\Delta\chi^2_{125} \le 6.18 $, where $\Delta \chi^2_{125} = {\chi^{2}_{125}}^{GM} - {\chi^{2}_{125}}^{SM}$, that is to say, the particular BP falls within the $2\sigma$ limit of the SM value \cite{95gev1}. 
Among these constraints, the bound coming from $H_3^+ \rightarrow \tau^+ \nu_{\tau}$\cite{h3pdecay} appears to be especially stringent when the $m_3$ lies in the range $140-160$ GeV. In addition, confirmatory checks for the constraints on $\sigma(p p \rightarrow H)\times BR(H \rightarrow \gamma \gamma)$ have also been carried out using the database available at HEPData \cite{hepdata}.  The constraint coming from the search of spin 0 particle in the diphoton final state at the ATLAS detector is the most stringent in the considered parameter space \cite{atlsdiph}, while the Yukawa coupling of $H$ was mainly constrained by the search in ref. \cite{htaublike}. 
The constraints coming from searches for the doubly charged scalar are also included but these do not pose any new constraint over and above the already existing ones \cite{lowdoubatls} \cite{lowdoubcms}, since the mass range of doubly charged scalar lies in the range $90 - 200$ GeV for our desired diphoton signal strength.

\vspace{0.15cm}

Table~\ref{sgbp} lists our BPs for signal estimation, consistent with the constraints, mentioned above. Points in the GM parameter space, leading to signals detectable at the high-luminosity LHC (HL-LHC), have $m_H$ lying in the approximate range of 160-250 GeV. We have chosen 6 benchmark points in this range to illustrate our results.

\begin{center}
	\begin{table}
		\hspace{1cm}
		\begin{tabular}{| c c c c | } 
			\hline
			\rule{0pt}{3ex}
			\hspace{0.5 cm} BP \hspace{0.5 cm} & $m_H$[GeV] \hspace{0.5 cm} & BR$(H \rightarrow \gamma \gamma)$ \hspace{0.5 cm} & $\sigma(pp\rightarrow H)_{ggF} \times BR(H \rightarrow \gamma \gamma)$ [fb]\\ [0.5ex] 
			\hline
			\rule{0pt}{3ex}
			BP1 & 160 & 0.015 & 9.52 \\ 
			\rule{0pt}{3ex}
			BP2 & 180 & 0.009 & 5.28 \\
			\rule{0pt}{3ex}
			BP3 & 200 & 0.011 & 4.95 \\
			\rule{0pt}{3ex}
			BP4 & 220 & 0.008 & 4.84 \\
			\rule{0pt}{3ex}
			BP5 & 240 & 0.008 & 3.40 \\ 
			\rule{0pt}{3ex}
			BP6 & 246 & 0.073 & 2.55 \\ [1ex] 
			\hline
		\end{tabular}
		\caption{Selected benchmark points, the corresponding diphoton branching ratio of $H$ and values of the highest leading order $\sigma(pp\rightarrow H)_{ggF}\times BR(H \rightarrow \gamma \gamma)$  as obtained from the scan. These values are obtained after applying the generation level cuts of $p_T^\gamma \ge 40$~ GeV , $|\eta| \le 2.7$ and $120 \ \rm GeV \le m_{\gamma\gamma} \le  280 \ \rm GeV$. }
		\label{sgbp}
	\end{table}
\end{center}
\vspace{-1.5cm}

\section{Simulation of signal and backgrounds}
\label{sigbkg}
It is already known that due to relatively less hadronic activity, the diphoton channel emerges as one of the cleanest processes for Higgs signals. Since part of the decay is mediated through a top quark loop, this decay mode appears to be most useful for a low mass Higgs whose other tree-level decay modes are suppressed. However due to the reasons explained in section \ref{model}, in the GM model this mode turns out to be significant in a relatively higher mass range of 160-250 GeV. Due to the high gluon flux at the LHC, we consider gluon fusion to be the production mode of the signal Higgs. The signal rate at leading order (LO) is tabulated in Table~\ref{sgbp}, following the guideline set down in section \ref{bp}. 
\vspace{0.1cm}

In all the benchmarks, the signal events have been generated with   MadGraph5\_\ aMC@NLO   \cite{mg5}, with a cut of $p_T \ge 40 ~ \rm GeV$  and $|\eta| \le 2.7$ on both the photons and $120 ~ \rm GeV \le m_{\gamma \gamma} \le 280 ~ \rm GeV$. To account for NLO contribution, the cross-section is multiplied by a $k$-factor \cite{nlo}. The model file for the signal is generated with Feynrules\cite{fr}. 
\vspace{0.1cm}

Despite being one of the most sensitive channels for the discovery of a new scalar $H$, the signal process can be mimicked by quite a few SM processes which constitute non-negligible backgrounds. These consist mostly of the diphoton irreducible background and the reducible backgrounds arising from jets faking as photons (j$\gamma$ and jj).
\begin{itemize}
\item The irreducible background consists of two isolated photons emerging from $q\bar{q}$ annihilation as well as from $g g \rightarrow \gamma \gamma$ through a one loop box diagram. The former has been generated in the next-to-leading order (NLO) using   MadGraph5\_\ aMC@NLO   \footnote{The background from  $g g \rightarrow \gamma \gamma$ is at $\mathcal{O}(\alpha_s \alpha)$ i.e the leading order itself.} . A generation level cut of $p_T \ge 40 ~ \rm GeV$  and $|\eta| \le 2.7$ has been applied on each photon. Additionally, an invariant mass cut of $120 ~ \rm GeV \le m_{\gamma \gamma} \le 280 ~ \rm GeV$ has been imposed on the photon pair.
\item Non-prompt photons within hadronic jets, misidentified as prompt photons when a major fraction of the jet energy is carried by the photons,  are the main source of reducible background. A jet, where $\pi^{0}$, $\eta$, or $\rho$ particles carry a large fraction of momentum of the jet and subsequently decay into two closely aligned photons, lead to the deposition of energy within ECAL (Electromagnetic Calorimeter) and effectively mimics the signature of a lone photon. As indicated in Table~\ref{cut}, these backgrounds are largely suppressed by the isolation and $p_T$ cuts. 
\item The reducible backgrounds $(j\gamma , jj)$ have been generated with Pythia8 \cite{pythia} in the leading order (LO) since they require an exorbitant amount of statistics.   As demonstrated in \cite{dijetnlo}, this leads to a conservative estimate of the signal significance, even when NLO effects are included. From the $jj$ sample, very few events passed the final selection cuts. We modeled the $jj$-background with an exponential and estimated its parameters by unbinned likelihood fit to the surviving events. We generated a histogram corresponding to $3000~ fb^{-1}$ by resampling from this exponential distribution.
\item All the hard processes corresponding to signal and backgrounds (except $q\bar{q} \rightarrow \gamma \gamma$) 
have been generated with the parton density function NNPDF23LO \cite{nnpdf}. The corresponding NLO version has been used for the $q\bar{q}$ annihilation process. The renormalization and factorization scales have been set to their default values namely $m_Z$.

\item In order to estimate reducible backgrounds as effectively as possible, we first produce an EM-enriched sample following the procedure described below, which we have closely adapted from \cite{bhowmik}. By EM-enriched sample, we mean that the events are selected such that the jet in the events are rich in $\pi^0, \eta,  \ or  \ \rho$ meson, which have a large diphoton branching fraction and hence can fake the signal with a high probability. This sample then undergoes a detector simulation through Delphes-3.5.0 \cite{delphes} .

\item The jets have been clustered by the Anti-kT \cite{antikt} algorithm using Fastjet \cite{fastjet} built within Delphes. After running the detector simulation, acceptance cuts of $p_T^{\gamma} \ge 10 ~ \rm GeV$ and $|\eta| \le 2.5$ have been applied on the photons. 
\end{itemize}

\textit {EM enrichment:} The faking backgrounds mainly originate in the QCD multijet processes $p p \rightarrow j \gamma$ and $p p \rightarrow j j$. Though the faking probability is $10^{-4}$ and $10^{-7}$ respectively for $j \gamma$ and $j j$ backgrounds, as revealed from our Monte Carlo studies, the sheer enormity of their cross sections ($10^3 ~ pb$ and $10^7 ~ pb $ within the generation level cuts), makes them non-negligible. The hard processes have been generated with $p_T \ge 40 \ \rm GeV$, $m_{j \gamma} \ m_{jj} \ge 120 \ \rm GeV$, and $|\eta| \le 2.7$. After showering, all particles with $p_T \ge 5 \ \rm GeV$ and $|\eta| < 2.7$ are collected in a list and we call these particles as `seed'. We made sure that the seed had the highest $p_T$ within a radius of $\Delta R = 0.09$ around it. Photon candidates are then formed by adding the $p_T$ and energies of all the particles within $\Delta R = 0.09$ to those of the seed itself. Events with two photon candidates with $p_T \ge 40 ~ \rm GeV$ are selected, as these events will have a higher probability of being fake. Since these backgrounds need to be generated with high statistics, we also applied an isolation criterion on the photon candidates, that is, the scalar $p_T$ sum of all the particles within radius $0.02 \le \Delta R \le 0.4$ around the photon candidate, is required to be smaller than $12 \%$ of the $p_T$ of the photon candidate. 
\vspace{0.2 cm}

\begin{figure}[htb]
 \begin{subfigure}[b]{0.45\textwidth}
         \centering
         \includegraphics[width=\textwidth]{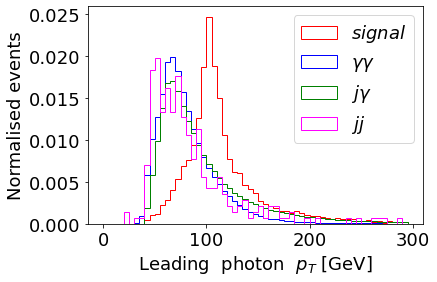}
         \caption{}
         \label{leadpt}
     \end{subfigure}
     \hfill
     \begin{subfigure}[b]{0.45\textwidth}
         \centering
         \includegraphics[width=\textwidth]{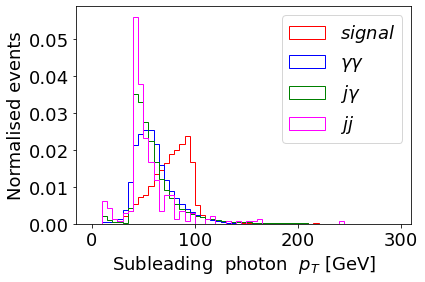}
         \caption{}
         \label{subleadpt}
     \end{subfigure}
\caption{{Left: Leading photon $p_T$ distribution of signal and backgrounds, Right: Sub-leading photon $p_T$ distribution of signal and backgrounds. Both distributions are shown for BP3}}
\label{pt}
\end{figure}

\begin{figure}[htb]
\centering
\includegraphics[width=0.55\linewidth]{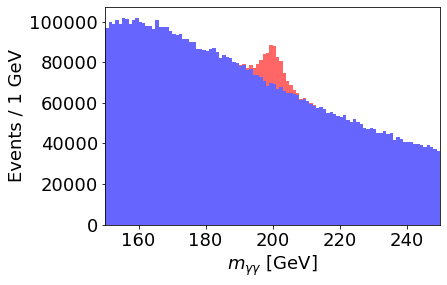}
\caption{Invariant mass distribution (unnormalized, corresponding to $\int \mathcal{L}dt = 3000 fb~^{-1}$) corresponding to BP3. The signal has been scaled up by a factor of 15. Events coming from background marked in blue and excess of events due to signal shown in red.}
\label{mass}
\end{figure}
\vspace{0.2cm}

 The distributions of leading and subleading photon $p_T$ for BP3 have been shown in Fig. \ref{pt} and the distribution in invariant mass of the diphoton system, corresponding to $\int \mathcal{L}dt = 3000 fb^{-1}$ is shown in Fig. \ref{mass}. From Fig. \ref{pt}, we can see that all the backgrounds lie mainly below 80 GeV for the leading and 70 GeV for the subleading photon $p_T$ while the signal peaks at around 100 GeV for the leading and around 90 GeV for the subleading photon $p_T$. For the benchmark point BP3, we estimate the significance of the predicted signal to be 7.1 with 3000 $fb^{-1}$ of data in the cut-based analysis. A typical signal shape in this region is shown in Fig.\ref{mass}. We have multiplied the signal strength by a factor of 15 for visual clarity. The sidebands chosen for this mass peak are 170-195 GeV and 205-230 GeV, which we used for calculating the significance (please see section \ref{llr} for details).  These considerations have gone into the choice of cuts applied, which along with the cut-flow table are adumbrated in Table~\ref{cut}. 

  As has been mentioned above, we have used an isolation criterion significantly looser than the tight photon identification criteria used in LHC experiments in Run $I$ and $II$ (e.g. Ref \cite{atlsdiph}, \cite{ATLAS_hga2}, \cite{CMS_hga}). Our isolation criterion demands the scalar $p_T$ sum of all the particles within the isolation cone to be smaller than $12 \%$ of the $p_T$ of the photon candidate. A tighter isolation criterion would have completely killed the dijet background in the $jj$ sample we could produce within our available computational resources and could potentially lead to an overestimation of signal significance. This makes our analysis conservative. With $j\gamma$ and $jj$ background suppression similar to that in ref \cite{atlsdiph}, the discovery significance can be about 30\% better than the one we have got based on a cut and count analysis. But in our optimized analysis utilizing a neural network with the isolation variable as one of its inputs, the suppression of the reducible background with respect to the irreducible background is comparable to that reported in ref. \cite{atlsdiph}.
 
\begin{table}
\begin{tabular}{| c c c c c c | } 
 \hline
 \rule{0pt}{3ex} 
 $m_H$[GeV] \hspace{0.5 cm} & cuts applied \hspace{0.5 cm} & signal [fb] \hspace{0.5 cm} & $\gamma \gamma$ [pb] \hspace{0.5 cm} & $j \gamma$ [pb] \hspace {0.5 cm} & jj [pb]  \\ [0.5ex] 
 \hline
 
 \rule{0pt}{3ex}

 200 &  initial cross-section   &  4.95   &   9.17  &  $4.89 \times 10^3$  &  $2.01 \times 10^7$ \\
 
      \rule{0pt}{3ex}
      
      &  acceptance   &  3.17   &   6.08  &  3.97  & 2.87 \\
     
      \rule{0pt}{3ex}
      
     &  $p_T^{l} \ge 70 \ \rm GeV \ , \ p_T^{sl} \ge 60 \ \rm GeV $   &  2.72   &   2.02  &  1.35  & 0.64 \\
     
      \rule{0pt}{3ex}
      
     &  $170 \ \rm GeV \le m_{\gamma \gamma} \le 230 \ \rm GeV $ &  2.72   &   0.79  &  0.46  & 0.08 \\
 \hline
\end{tabular}
\caption{The cut flow table corresponding to BP3. $p_T^l (p_T^{sl})$ corresponds to the leading (subleading) photon $p_T$.}
\label{cut}
\end{table}

\section{Results}
\label{rslt}
\subsection{\large{Significance from likelihood ratio method}}
\label{llr}
We have used, {\it mutatis mutandis}, the profile likelihood ratio method for binned data as formulated in reference \cite{Cowan} to calculate signal significance. The statistical analysis is designed to look for a local excess in $m_{\gamma\gamma}$ distribution. In this study, we have used $m_{\gamma\gamma}$ distribution with a bin size of 1 GeV, for signal and each background.
The number of events in each bin is assumed to be a Poisson distribution. We model the signal with a Gaussian centered at $m_H$. The mean signal yield $s_i$ at $i^{th}$ bin is expressed as,
\begin{equation}
s_i = a_1 e^{-\frac{(x_i - m_H)^2}{2\sigma^2}}
\end{equation}
$a_1$ and $\sigma$ were determined from the $m_{\gamma \gamma}$ distribution corresponding to the signal, by maximizing the likelihood. As found from the fit, 3 times the standard deviation of the Gaussian peak ($3\sigma) \approx 10 ~\rm GeV$ for BP2-BP5, while for BP1 $3\sigma \approx 5 ~\rm GeV$. Hence for BP1, $(m_H \pm 5) ~ \rm GeV$ and for the rest, $(m_H \pm 10) ~ \rm GeV$ have been defined as the signal region.
The mean entry from the background was estimated by blinding the signal region and fitting the sideband with an exponential function.  Hence, the expected event rate $\lambda_{b_i}$ in the $i^{th}$ bin from the background is given by,
\begin{equation}
\lambda_{b_i} = a_0 e^{-b x_i}
\end{equation}
where $x_i$ is the bin center. Each type of background was modeled separately in the same way and the shape parameter $b$ was extracted from the fit while $a_0$ was determined using the number of events expected at $3000 ~ fb^{-1}$. 

With these, the expected number of events, $\lambda_i$ in the $i^{th}$ bin is given by,
\begin{equation}
\lambda_i = a_0 e^{-b x_i} + \mu a_1 e^{-\frac{(x_i - m_H)^2}{2\sigma^2}}
\end{equation}
$a_1$ and $\sigma$ have been estimated from the signal. Except for the signal parameters i.e $m_H, a_1$, and $\sigma$, all other parameters $(a_0 , b, \mu)$ were allowed to vary freely in the fit. $\mu$ is the parameter of interest and $(a_0 , b)$ are the nuisance parameters. The normalizations $a_0$ and $a_1$ contain the cross-section, integrated luminosity, efficiency for background and signal respectively.

The likelihood function for the parameters $(a_0,b,\mu)$,
\begin{equation}
\mathcal{L}_(a_0,b,\mu) = \Pi_i \frac{\lambda_i^{n_i} e^{-\lambda_i}}{n_i!}
\end{equation}
Where, $n_i$ is the observed number of events in the $i^{th}$ bin and the product is over the bins.
The likelihood function for the parameters under the background only hypothesis is given by,
\begin{equation}
\mathcal{L}_1(a_0,b,\mu=0) = \Pi_i \frac{{\lambda_{b_i}}^{n_i} e^{-\lambda_{b_i}}}{n_i!}
\end{equation}
$a_0$ and $b$ are evaluated by minimizing $-log(\mathcal{L}_1)$. Finally, the test statistic is given by,
\begin{eqnarray}
\label{sys0}
q_0 &=& -2log\frac{\mathcal{L}_1(\hat{\hat{a_0}},\hat{\hat{b}},\mu=0)}{\mathcal{L}_2(\hat{a_0},\hat{b},\hat{\mu})}  \ \ \ \ \hat{\mu} \ge 0 \nonumber \\
   &=& 0  \ \ \ \ \hat{\mu} < 0   
\end{eqnarray}

In the denominator, ${\mathcal{L}_2(\hat{a_0},\hat{b},\hat{\mu})}$ denotes the global maximum of the likelihood function and $(\hat{a_0},\hat{b},\hat{\mu})$ are the values of $a_0, \ b$, and $\mu$ that maximize the likelihood globally. In the numerator, ${\mathcal{L}_1(\hat{\hat{a_0}},\hat{\hat{b}}}, \mu=0)$ denotes the maximum of the likelihood function for $\mu=0$, and $\hat{\hat{a_0}},\hat{\hat{b}}$ are the values of $a_0, \ b$ that maximize the likelihood function when $\mu = 0$. The significance is given by,
\begin{equation}
    Z = \sqrt{q_0}
\end{equation}
As shown in \cite{Cowan} in the asymptotic limit and when the signal $(s)$ is much smaller than background $(B)$, Z-score becomes,
\begin{equation}
    Z = \Sigma_i \frac{s_i}{\sqrt{B_i}}
\end{equation}
where the sum is over the number of bins in the histogram. This explains the improvement in significance with a $\frac{s}{B} \approx 2\%$, even in the presence of systematics.
\newline
%Here we note that $a_0$ and $b$ are nuisance parameters and to account for systematics, we take into account, the probability density functions (PDF) of these parameters. For this study we have considered the PDF corresponding to $a_0$ only. In presence of systematics, the test statistic is given by,

To account for systematics, we generated toy Monte-Carlo samples for each background following the fit function given by,
\begin{equation}
\lambda_{b_i} = (1+\Delta_s)a_0 e^{-bx_i}
\end{equation}
where $\Delta_s$ is the amount of systematic error. These backgrounds have been used to calculate significance in the presence of systematics, by following the above mentioned procedure.

\subsection{\large{Cut-based analysis: Parameter space with $\ge 3\sigma$ significance at $3000$~$fb^{-1}$}}
We present the regions of the parameter space of the GM scenario where the diphoton signal
is discernible at the level of $3\sigma$ or more for an integrated luminosity of $3000~fb^{-1}$
at the LHC. The calculation of signal significance has been carried out according to the algorithm
outlined in the previous subsection.

\begin{center}
\begin{table}[h]
\begin{tabular}{|c c c c c| } 
 \hline
 \rule{0pt}{3ex}
 \hspace{0.5 cm} BP \hspace{0.5 cm} & $cut1$ \hspace{0.5 cm} & $cut2$ & significance & significance \\ 
 \hspace{0.5 cm} & \hspace{0.5 cm} & \hspace{0.5 cm} & without systematics & with 10\% systematics \\[0.5ex] 
 \hline
 
 \rule{0pt}{3ex}
 
 BP1 & $p_T^l > 55 \ , \ p_T^{sl} > 45$ \ \ & \ \ $130 < m_{\gamma \gamma} < 180$ & 10 & 9.4 \\
 
\rule{0pt}{3ex}

 BP2 & $p_T^l > 65 \ , \ p_T^{sl} > 55$ \ \ & \ \ $150 < m_{\gamma \gamma} < 210$ & 5.7 & 5.3 \\
 
 \rule{0pt}{3ex}
 
 BP3 & $p_T^l > 70 \ , \ p_T^{sl} > 60$ \ \ & \ \ $170 < m_{\gamma \gamma} < 230$ & 7.6 & 7.1 \\

 \rule{0pt}{3ex}
 
 BP4 & $p_T^l > 80 \ , \ p_T^{sl} > 70$ \ \ & \ \ $190 < m_{\gamma \gamma} < 250$ & 10.4 & 10 \\
 
 \rule{0pt}{3ex}
 
 BP5 & $p_T^l > 90 \ , \ p_T^{sl} > 80$ \ \ & \ \ $210 < m_{\gamma \gamma} < 270$ & 9.1 & 8.7 \\  
  
 \rule{0pt}{3ex}
  
 BP6 & $p_T^l > 90 \ , \ p_T^{sl} > 80$ \ \ & \ \ $210 < m_{\gamma \gamma} < 270$ & 6.7 & 6.4 \\ [1ex]

\hline
 \end{tabular}
\caption{Analysis level cuts for each BP and the corresponding significance both without systematics and with 10\% systematics at 3000 $fb^{-1}$. $p_T^l (p_T^{sl})$ corresponds to the leading (subleading) photon $p_T$}
\label{tab-signi}
\end{table}
\end{center}
\vspace{-1.2cm}

\begin{figure}[htb!]
     %\centering
     \begin{subfigure}[h]{0.48\textwidth}
         \centering
         \includegraphics[width=1.1\textwidth,height=0.8\textwidth]{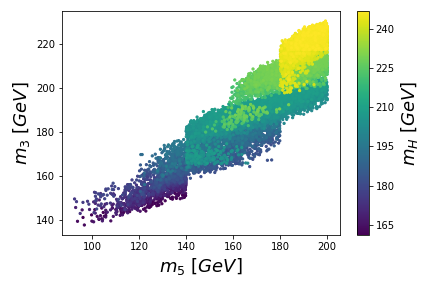}
         \caption{}
         \label{sys3000a}
     \end{subfigure}
     \hfill
     \begin{subfigure}[h]{0.48\textwidth}
         \centering
         \includegraphics[width=1.1\textwidth,height=0.8\textwidth]{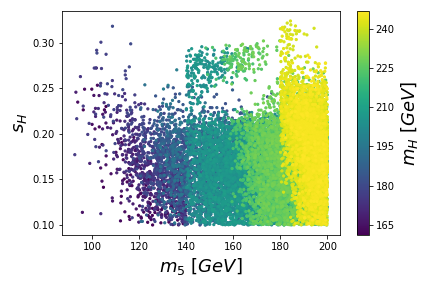}
         \caption{}
         \label{sys3000b}
     \end{subfigure}
%    \end{figure}%    
%     \begin{figure}[h!]\ContinuedFloat 
     %\hfill
     \begin{subfigure}[h]{0.48\textwidth}
         \centering
         \includegraphics[width=1.1\textwidth,height=0.8\textwidth]{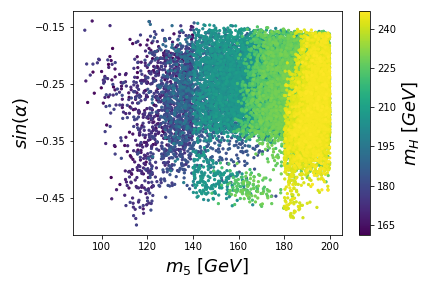}
         \caption{}
         \label{sys3000c}
     \end{subfigure}
     \hfill
     \begin{subfigure}[h]{0.48\textwidth}
         \centering
         \includegraphics[width=1.1\textwidth,height=0.8\textwidth]{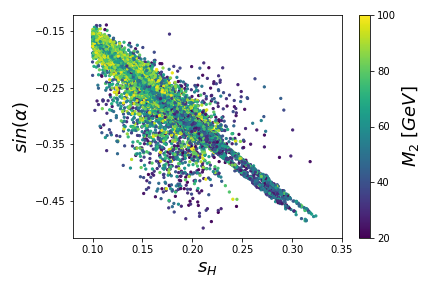}
         \caption{}
         \label{sys3000d}
     \end{subfigure}
     %\end{figure}
     %\begin{figure}[H]
        %\hfill
    %\end{figure}%    
    %\begin{figure}[h!]\ContinuedFloat    
    %\hspace*{3cm}    
     \begin{subfigure}[b]{0.48\textwidth}
         \centering
    \includegraphics[width=1.1\textwidth,height=0.8\textwidth]{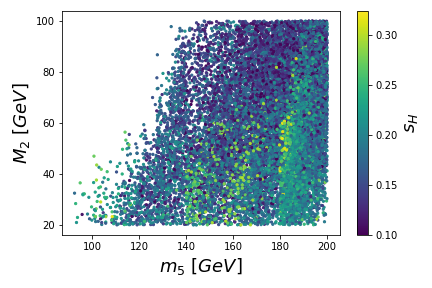}
         \caption{}
         \label{sys3000e}
     \end{subfigure}
        \caption{Regions of the parameter space (\ref{sys3000a} , \ref{sys3000b} , \ref{sys3000c} , \ref{sys3000d} in $m_5 - m_3$ , $m_5 - s_H$ , $m_5 - \sin \alpha$ , and $s_H - \sin \alpha$ planes respectively with $m_H$ color-coded, \ref{sys3000e} in $m_5 - M_2$ plane with $s_H$ color-coded) that can be probed with a significance $\ge$ 3$\sigma$ at the HL-LHC, with $\int{\mathcal{L}} dt$ = 3000 $fb^{-1}$ in presence of 10\% systematics.}
        \label{regsys3000}
\end{figure}

Table~\ref{tab-signi} shows the significance, corresponding to each BP, both with systematic uncertainty set to zero and with 10\% systematic uncertainty. In order to obtain Fig. \ref{regsys3000}, the parameter space has been divided into five parts corresponding to the first five BPs with $m_H$ within the upper 20 GeV band of the benchmark $m_H$. The background, as estimated for the BP, has been used to calculate the significance for the respective region. The signal cross-section has been obtained using   MadGraph5\_\ aMC@NLO   for each point in the parameter space, and the efficiency factor of the BP has been multiplied to account for the cut efficiency.

\vspace{0.2cm}

Since no significant change happens in terms of parameter space upon inclusion of systematics, here we present the parameter regions corresponding to a significance of $\ge 3 \sigma$ at 3000 $fb^{-1}$ in the presence of 10\% systematic uncertainty. We make the following observations based on Fig.~\ref{regsys3000}.

\begin{itemize}
\item The predicted diphoton signals are expected to be significant in the region with $m_H$ approximately
in the range 160 - 250 $\rm GeV$, as seen in Fig. \ref{sys3000a}.

\item The same figure also shows that the signal is favoured mostly for $m_5 \le m_3$. Even in the limited number of cases where this order is reversed,  the mass splitting is mostly within 10-30 GeV. This suppresses the decay $H \rightarrow H^{\pm}_3 W^{\mp}$, thus enhancing the diphoton decay branching ratio.

\item As is also seen in Fig. \ref{sys3000b},  $m_5 \le 200$ GeV  favours the signal.
This is also because the loop contributions are enhanced for lower masses of the
doubly charged scalar. Additionally, $H^{\pm}_3$ also contributes significantly whenever the corresponding trilinear coupling
is substantial.

\item A substantial portion of the low $m_5$ region, which yields significant signal strength, also corresponds to a relatively large $s_H$ (read the triplet contribution to the $W/Z$ mass). This happens even when $H_5^{\pm\pm}$ decays exclusively to $W^{\pm}W^{\pm}$. The enhancement in $s_H$ happens due to the fact that the low mass $H_5^{\pm\pm}$ has a suppressed branching ratio to $W^{\pm}W^{\pm}$ and therefore the LHC bounds are more relaxed. Low $m_H$ also favours this, as the enhanced production rates due to large $s_H$ and hence large $\zeta_F$ lead to fermionic signals that are liable to be swamped by backgrounds.

\item The favoured regions indicated in the figures correspond to the $\Delta L = 2$ Yukawa coupling  $Y_{ll}$, being $\le 10^{-6}$. For larger values of $Y_{ll}$, the absence of same-sign dilepton signal puts stringent lower limits on $m_5$ \cite{Primulando}, which causes suppression of the loop amplitudes leading to the diphoton signal.

\item As all the figures in Fig.~\ref{regsys3000} indicate, the signal is favoured for relatively low $m_5$ which is still allowed by all LHC-based studies reported so far. In this sense, the diphoton signals predicted in our analysis could be crucial for probing the low-$m_5$ regions in the GM parameter space, particularly for $m_5 \le 200$~GeV, where the searches based on $H_5^{\pm\pm}$ production do not work so well. This is true even when the search is carried out with $\int\mathcal{L} dt = 300 ~fb^{-1}$.

\item Figures \ref{sys3000c} and \ref{sys3000d} illustrate the signal regions in terms of $\sin\alpha$ as defined by equation \eqref{matrix}, which
is the measure of the doublet content of $H$, a feature that decides its production rate
in the gluon fusion channel. We find that the diphoton signal undergoes enhancement for large $|\sin\alpha|$, precisely due to the reason mentioned above.

\item It is also seen from Figures \ref{sys3000b}, \ref{sys3000d}, and \ref{sys3000e} that the diphoton channel serves as a sensitive probe even for small $s_H$. This is significant because searches based on the 5-plet suffer from very low signal strength, as both their production and decay processes are driven solely by $s_H$.

\item Fig. \ref{sys3000e} demonstrates that substantial signal strength can still be expected for small $M_2$ when $s_H$ is high. A high $s_H$ corresponds to a high $|\sin\alpha|$, as shown in Fig. \ref{sys3000d}, which enhances the doublet component in $H$ and thus its production rate via gluon fusion.
\end{itemize}

\vspace{0.2 cm}

\begin{figure}[htb!]
     %\centering
     \begin{subfigure}[h]{0.48\textwidth}
         \centering
\includegraphics[width=\textwidth,height=0.5\textwidth]{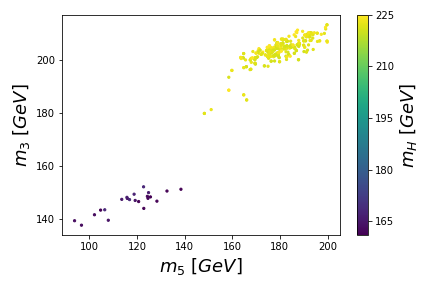}
         \caption{}
         \label{sys300a_3sig}
     \end{subfigure}
     \hfill
     %\centering
     \begin{subfigure}[h]{0.48\textwidth}
         \centering
         \includegraphics[width=\textwidth,height=0.5\textwidth]{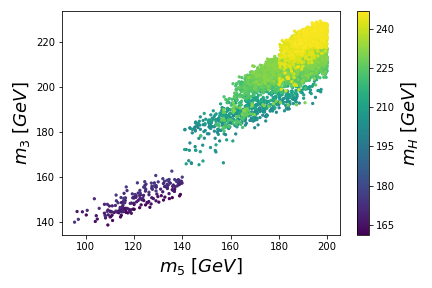}
         \caption{}
         \label{sys300a_2sig}
     \end{subfigure}
     \begin{subfigure}[h]{0.48\textwidth}
         \centering
\includegraphics[width=\textwidth,height=0.5\textwidth]{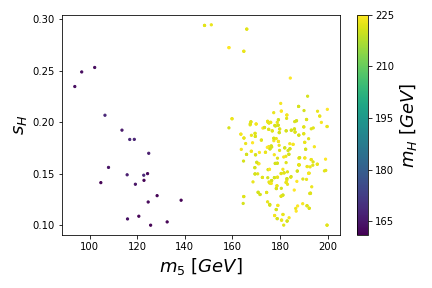}
         \caption{}
         \label{sys300b_3sig}
     \end{subfigure}
     \hfill
     \begin{subfigure}[h]{0.48\textwidth}
         \centering
\includegraphics[width=\textwidth,height=0.5\textwidth]{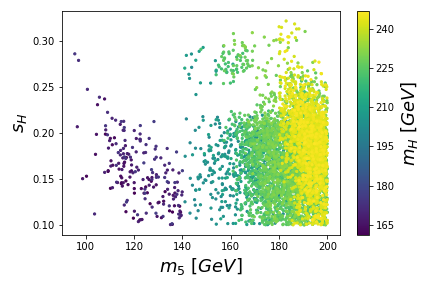}
         \caption{}
         \label{sys300b_2sig}
     \end{subfigure}
\hfill
     \begin{subfigure}[h]{0.48\textwidth}
         \centering
\includegraphics[width=\textwidth,height=0.5\textwidth]{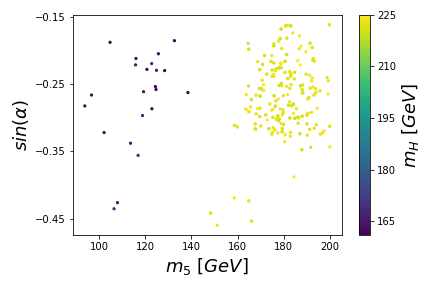}
         \caption{}
         \label{sys300c_3sig}
     \end{subfigure}
     \hfill
     \begin{subfigure}[h]{0.48\textwidth}
         \centering
\includegraphics[width=\textwidth,height=0.5\textwidth]{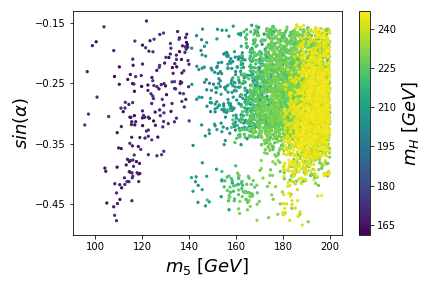}
         \caption{}
         \label{sys300c_2sig}
     \end{subfigure}
   \begin{subfigure}[h]{0.48\textwidth}
         \centering
\includegraphics[width=\textwidth,height=0.5\textwidth]{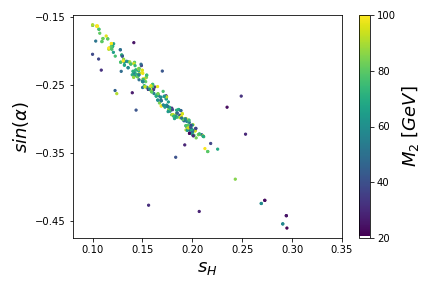}
         \caption{}
         \label{sys300d_3sig}
     \end{subfigure}
        \hfill
        \begin{subfigure}[h]{0.48\textwidth}
         \centering
\includegraphics[width=\textwidth,height=0.5\textwidth]{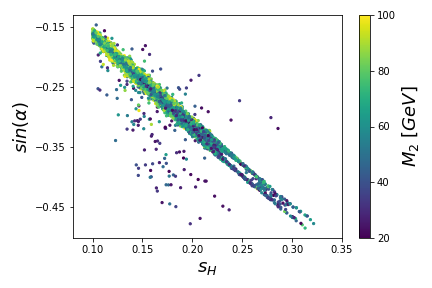}
         \caption{}
         \label{sys300d_2sig}
     \end{subfigure}
       \hfill
     \begin{subfigure}[h]{0.48\textwidth}
     \centering
 \includegraphics[width=\textwidth,height=0.5\textwidth]{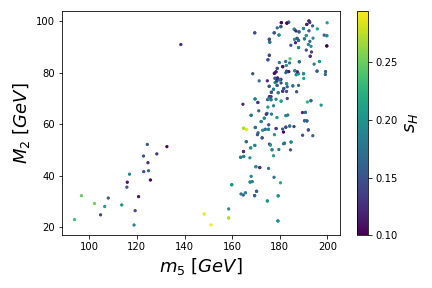}
         \caption{}
         \label{sys300e_3sig}
     \end{subfigure}
     \hfill
     \begin{subfigure}[h]{0.48\textwidth}
         \centering
    \includegraphics[width=\textwidth,height=0.5\textwidth]{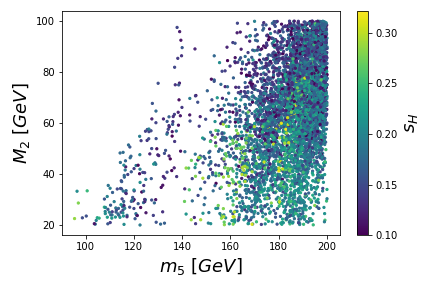}
         \caption{}
         \label{sys300e_2sig}
     \end{subfigure}
        \caption{Left column: \ref{sys300a_3sig} , \ref{sys300b_3sig} , \ref{sys300c_3sig} , \ref{sys300d_3sig} , \ref{sys300e_3sig} same as in Fig. \ref{regsys3000} but with $\int{\mathcal{L}} dt$ = 300 $fb^{-1}$. Right column: Regions (\ref{sys300a_2sig} , \ref{sys300b_2sig} , \ref{sys300c_2sig} and \ref{sys300d_2sig} in $m_5 - m_3$ , $m_5 - s_H$ , $m_5 - \sin \alpha$ , and $s_H - \sin \alpha$ planes respectively with $m_H$ color-coded,  \ref{sys300e_2sig} in $m_5 - M_2$ plane with $s_H$ color-coded) that can be probed at 2$\sigma$ level in presence of 10\% systematics with $\int{\mathcal{L}} dt$ = 300 $fb^{-1}$.}
        \label{regsys300}
\end{figure}
\afterpage{\clearpage }
\vspace{-0.5cm}

Since the potential for covering the GM parameter space via diphoton signals appears quite promising, it is useful to assess the extent of this coverage even before the HL-LHC begins operation. To address this, we present in Fig. \ref{regsys300} the regions of parameter space that can be probed via diphoton signals with an integrated luminosity of $\int \mathcal{L} dt = 300~fb^{-1}$. This figure demonstrates that the diphoton signal becomes significant even before the conclusion of Run-3 and will certainly be notable in the early phase of the HL-LHC. For comparison, we also include regions that can be probed at the 2$\sigma$ level with the same luminosity.

\subsection{\large{Results based on Neural Network}}
	Neural network-based techniques are well known for their ability to exploit the information in the data to optimally distinguish between the signal and background processes. Essentially, the neural networks are universal function approximators that can shatter the input space with much more flexible surfaces than the orthogonal set of planes used in a typical cut-based analysis. In this section, we describe an Artificial Neural Network (ANN) \cite{ann} that we have built for signal-background separation using photon isolation and kinematic variables. We have used the toolkit Keras \cite{keras} with TensorFlow \cite{tf} as the backend for the implementation of ANN. The network used in this analysis is composed of 5 hidden layers with 100 nodes each. A dropout of 20\% has been used for regularization. The $tanh$ function has been used as the activation function  for the input layer and the first two hidden layers, while $ReLU$ was used for the remaining layers, except for the output one. Since this is a binary classification problem, $sigmoid$ has been used as the activation function for the output layer. Binary crossentropy was used as the loss function, with Adam being the optimizer. Training has been done over 50 epochs with a batch size of 640 for each. A cut of $|\eta| \le 2.5$ has been applied to each photon before passing an event to the network. Also, events having diphoton invariant mass in the window $(m_H - 30)~ \rm GeV \le m_{\gamma \gamma} \le (m_H + 30)~ \rm GeV$ have been used in the network for both training and testing purposes. We have used the following variables to train the network.
	\vspace{0.4cm}
	
	1. The scalar sum of $p_T$ in an isolation cone of $\Delta R \le 0.5$, relative to the $p_T$ of the candidate photon \newline
	2. $\frac{p_T^{l}}{m_{\gamma \gamma}}$ and $\frac{p_T^{sl}}{m_{\gamma \gamma}}$, where $p_T^{l} (p_T^{sl})$ is the leading (subleading) photon $p_T$. \newline
	3. $\frac{E^{l}}{m_{\gamma \gamma}}$ and $\frac{E^{sl}}{m_{\gamma \gamma}}$, where $E^{l} (E^{sl})$ is the leading (subleading) photon energy. \newline
	4. $\Delta|\eta|$ and $\Delta|\phi|$ between two photons. \newline
	5. space opening angle $\Delta \theta_{12}$ between the two photons.
	
\begin{figure}[h]
		\centering
    \includegraphics[width=0.8\linewidth,height=0.6\linewidth]{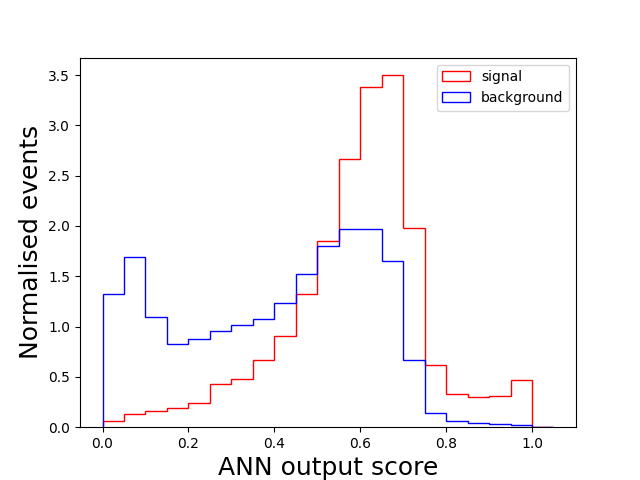}
		\caption{The neural network output score distribution}
		\label{dnn score}
	\end{figure}

	The network output distribution for signal and background for BP3 is shown in Fig. \ref{dnn score}. We observe a 68\% improvement in the area under the curve relative to the cut-based analysis. This translates to an improvement in the significance of discovery from 7.1 to 13.7 in Z-score for this benchmark point. The luminosity required to achieve a significance of 5$\sigma$ is reduced from 1500 $fb^{-1}$ to 399 $fb^{-1}$.  The improvements for all benchmark points are listed in Table~\ref{tab-signi-ann}. 
	\vspace{0.2cm}
	
\begin{figure}[h!]
	\centering
	\begin{subfigure}[b]{0.48\textwidth}
		\centering
	\includegraphics[width=1.1\textwidth,height=0.8\textwidth]{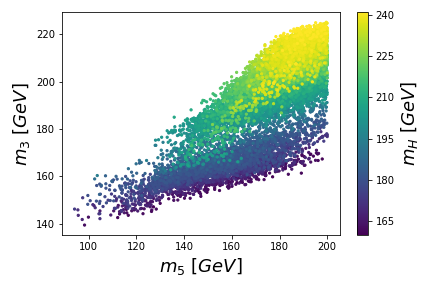}
		\caption{}
		\label{annsys3000a}
	\end{subfigure}
	\hfill
	\begin{subfigure}[b]{0.48\textwidth}
		\centering
	\includegraphics[width=1.1\textwidth,height=0.8\textwidth]{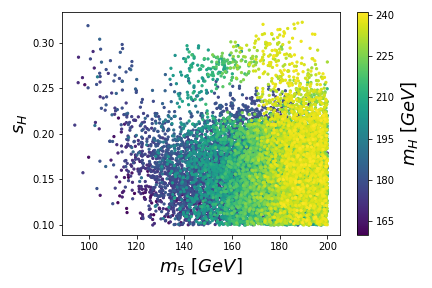}
		\caption{}
		\label{annsys3000b}
	\end{subfigure}
	\hfill
	\begin{subfigure}[b]{0.48\textwidth}
		\centering
	\includegraphics[width=1.1\textwidth,height=0.8\textwidth]{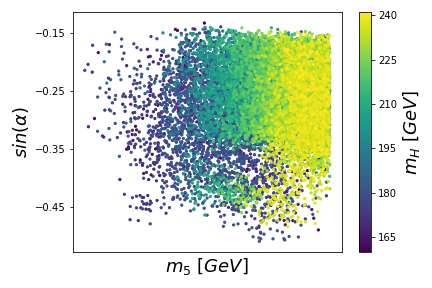}
		\caption{}
		\label{annsys3000c}
	\end{subfigure}
	\hfill
	\begin{subfigure}[b]{0.48\textwidth}
		\centering
	\includegraphics[width=1.1\textwidth,height=0.8\textwidth]{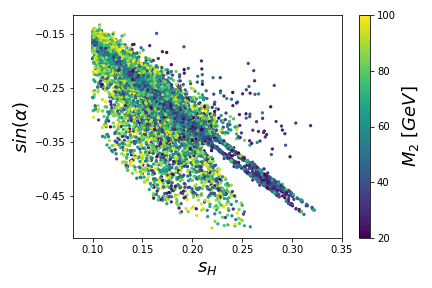}
		\caption{}
		\label{annsys3000d}
	\end{subfigure}
%\end{figure}% 
%\hspace{4cm} 
%\begin{figure}[h!]\ContinuedFloat     
	\begin{subfigure}[htb]{0.5\textwidth}
	%\centering
		\hspace{4cm}
\includegraphics[width=1.1\textwidth,height=0.8\textwidth]{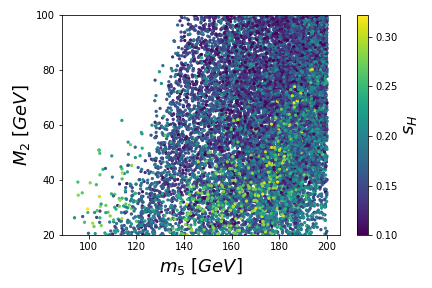}
		\caption{}
		\label{annsys3000e}
	\end{subfigure}
	\caption{Regions of the parameter space (\ref{annsys3000a},\ref{annsys3000b},\ref{annsys3000c},\ref{annsys3000d} in $m_5 - m_3$ , $m_5 - s_H$ , $m_5 - \sin \alpha$ , and $s_H - \sin \alpha$ planes respectively with $m_H$ color-coded, \ref{annsys3000e} in $m_5 - M_2$ plane with $s_H$ color-coded) that can be probed with a significance $\ge$ 3$\sigma$ at the HL-LHC, with $\int{\mathcal{L}} dt$ = 3000 $fb^{-1}$ in presence of 10\% systematics using ANN.}
	\label{regsysann3000}
\end{figure}

	Fig. \ref{regsysann3000} shows the regions of parameter space that become accessible to the HL-LHC at 3000~$fb^{-1}$ in the presence of 10\% systematics using ANN. A comparison between Fig.\ref{regsys3000} and Fig.\ref{regsysann3000} reveals that an appreciable portion of parameter space with $m_5 > m_3$ can be probed via this diphoton channel by implementing an ANN. Another notable change appears in terms of the parameter region that can be probed at 3$\sigma$ significance in Run-3 itself at 300~$fb^{-1}$. 
	Fig. \ref{regsysann300} clearly depicts that not only does the overall accessible parameter space increase, but also the mass range of 180-220 GeV, which was completely out of reach in the cut-based analysis, opens up using ANN for an integrated luminosity of 300~$fb^{-1}$.

\begin{center}
	\begin{table}[h]
		\centering
		\begin{tabular}{|c c c c c | } 
			\hline
			\rule{0pt}{3ex}
			
			\hspace{0.2 cm} BP \hspace{0.5 cm} & Significance (Cut-based) & $\int \mathcal{L} dt ~ [fb^{-1}]$ for  & significance (ANN)  & $\int \mathcal{L} dt ~ [fb^{-1}]$ for \\ 
			\hspace{0.5 cm} & with 10\% systematics & 5$\sigma$ (cut based)& with 10\% systematics   & 5$\sigma$ (ANN)\\[0.5ex] 
			\hline
			\rule{0pt}{3ex}
			
			BP1 & 9.4  & 849 & 17.4 & 248  \\
			
			\rule{0pt}{3ex}
			
			BP2 & 5.3 & 2670 & 15.6 & 308 \\
			
			\rule{0pt}{3ex}
			
			BP3 & 7.1 & 1488 & 13.7 & 399 \\
			
			\rule{0pt}{3ex}
			
			BP4 & 10 & 750 & 15.9 & 297\\
			
			\rule{0pt}{3ex}
			
			BP5 & 8.7 & 990 & 13.6 & 405 \\ 
			
			\rule{0pt}{3ex}
			
			BP6 & 6.8 & 1622 & 9.9 & 750 \\ [1ex] 
			
			\hline
		\end{tabular}
		\caption{Significance as obtained from both cut-based approach and ANN at 3000 $fb^{-1}$ upon inclusion of 10\% systematics. Required luminosity to reach 5$\sigma$ has also been noted.}
		\label{tab-signi-ann}
	\end{table}
\end{center}

\begin{figure}[htb!]
	%\centering
	\begin{subfigure}[b]{0.48\textwidth}
		\centering
		\includegraphics[width=1.1\textwidth,height=0.7\textwidth]{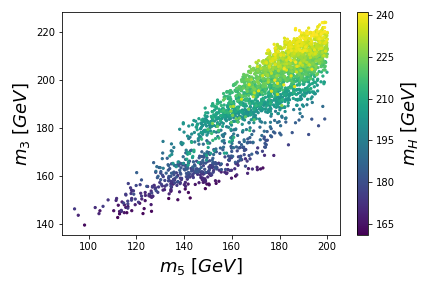}
		\caption{}
		\label{annsys300a}
	\end{subfigure}
	\hfill
	\begin{subfigure}[b]{0.48\textwidth}
		\centering
		\includegraphics[width=1.1\textwidth,height=0.7\textwidth]{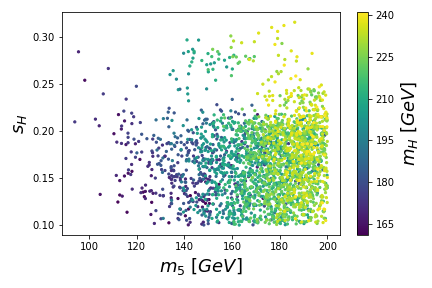}
		\caption{}
		\label{annsys300b}
	\end{subfigure}
	%\end{figure}%    
	%\begin{figure}[h!]
	%\hfill
	\begin{subfigure}[b]{0.48\textwidth}
		\centering
		\includegraphics[width=1.1\textwidth,height=0.7\textwidth]{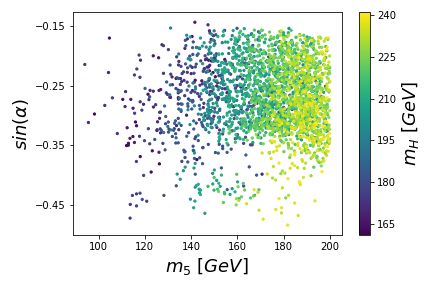}
		\caption{}
		\label{annsys300c}
	\end{subfigure}
	\hfill
	\begin{subfigure}[b]{0.48\textwidth}
		\centering
		\includegraphics[width=1.1\textwidth,height=0.7\textwidth]{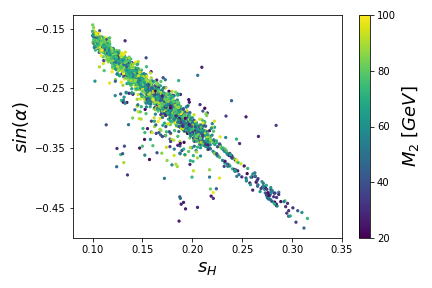}
		\caption{}
		\label{annsys300d}
	\end{subfigure}
\end{figure}% 
%\afterpage{\clearpage}   
\begin{figure}[h!]\ContinuedFloat    
	%\hspace{4cm} 
	\begin{subfigure}[b]{0.48\textwidth}
		\centering		\includegraphics[width=1.1\textwidth,height=0.7\textwidth]{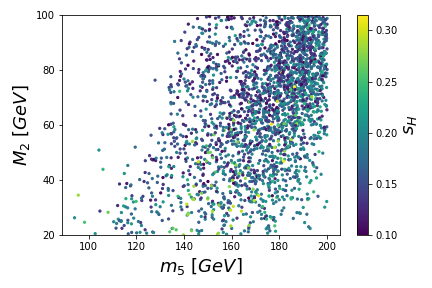}
		\caption{}
		\label{annsys300e}
	\end{subfigure}
	\caption{Regions of the parameter space (\ref{annsys300a},\ref{annsys300b},\ref{annsys300c},\ref{annsys300d} in $m_5 - m_3$ , $m_5 - s_H$ , $m_5 - \sin \alpha$ , and $s_H - \sin \alpha$ planes respectively with $m_H$ color-coded , \ref{annsys300e} in $m_5 - M_2$ plane with $s_H$ color-coded) that can be probed with a significance $\ge$ 3$\sigma$ at the LHC, with $\int{\mathcal{L}} dt$ = 300 $fb^{-1}$ in presence of 10\% systematics using ANN.}
	\label{regsysann300}
\end{figure}
\vspace{-1.5cm}

\section{Summary and conclusions}
\label{cncl}
We have identified the diphoton decay channel of the custodial SU(2) singlet scalar $H$ in the GM scenario as constituting a viable signal at the LHC. The main reasons for this
enhancement are (a) contributions of the doubly and singly-charged scalar loops, (b) suppression of the destructively interfering fermion
loops, (c) enhancement of the relevant trilinear scalar couplings in certain regions of the parameter space, and (d) suppression of the tree-level fermion and gauge boson pair decays due to the dominant $SU(2)_L$ triplet composition of $H$ as compared to  $h$. \newline
One should note that BR$(H \rightarrow \gamma \gamma)$ and thus the diphoton rate suffers from considerable suppression when the $H^{\pm\pm}$- induced loop diagram is absent. This can distinguish the GM scenario from, say, one with two Higgs doublets.\newline
It is generally held that the $W^\pm W^\pm$ decay channel for $H^{\pm\pm}$ is the best way of probing the GM scenario. However, this signal may be less appreciable when \newline
(a) $s_H$ is on the lower side\newline
(b) the $H^{\pm\pm}$ also has non-negligible branching ratios into $W^{\pm}H_3^{\mp}$ and $H_3^{\pm}H_3^{\mp}$. In such cases, the diphoton final state triggered by the triplet-dominated neutral state $H$ is found to be of considerable utility.\newline
The two-photon final state, with invariant mass peaking at $m_H$, thus constitutes an independent search channel for the GM model, whose importance in multichannel analyses hardly needs to be emphasized for a scenario with several free parameters.\newline 
The major backgrounds come from SM contributions to $\gamma\gamma, j\gamma$, and $jj$ production, all of which have been taken into account in our simulation.  The parameter space for
the GM model has been scanned over, ensuring consistency with data from the 125 GeV scalar, general constraints from extended scalar sector searches, along with theoretical limits
and indirect constraints coming from rare decays and precision electroweak measurements.\newline
The signal significance has been computed using the profile likelihood ratio method. It is
found that the regions most amenable to detection at the LHC are those corresponding to
$m_H$ in the approximate range 160-180 and 220-240 GeV. We have identified the regions
in the parameter space, for which the diphoton signal may have at least $3\sigma$ significance, with integrated luminosity of 3000 $fb^{-1}$ as well as 300 $fb^{-1}$. We have further shown that the explorable region expands significantly when one performs an analysis based on a neural network. These results show that the diphoton final state constitutes a valuable component of the search for the GM scenario, along with those final states that arise from the decays of the doubly charged scalar $H^{\pm\pm}$. We have shown in this study that it is possible to see a 5$\sigma$ excess in this channel even before the HL-LHC becomes operational. 

\section{Acknowledgement}
The work of RG is supported by a fellowship awarded by University Grants Commission, India. The authors acknowledge the support provided by the Kepler Computing facility, maintained by the Department of Physical Sciences, IISER Kolkata, for various computational needs. We also thank Debabrata Bhowmik, Suman Dasgupta, Shubham Dutta, Deep Ghosh, Jayita Lahiri, Jyotiska Panda, Sirshendu Samanta, Tousik Samui, Ritesh K Singh and Amir Subba for helpful discussions. The work of SB has been supported partially by a project sanctioned by the Department of Science and Technology, Government of India, entitled `Signatures of new physics at present and future Colliders'.
\vspace{0.5cm}
%\newpage

%\begin{figure}[H]
%\centering
%\includegraphics[width=0.55\linewidth]{m5-m3-mh-regplot-sys.png}
%\includegraphics[width=0.55\linewidth]{m5-sh-mh-regplot-sys.png}
%\includegraphics[width=0.55\linewidth]{m5-sa-mh-regplot-sys.png}
%\includegraphics[width=0.55\linewidth]{sa-sh-m2-regplot-sys.png}
%%\includegraphics[width=0.55\linewidth]{m5-m2-sh-regplot-sys.png}
%\caption{Figure 4(a) - 4(e)  show the regions of parameter space that can be probed with a significance $\ge$ 3$\sigma$ after including 10\% systematics at the HL-LHC.}
%\label{regsys}
%\end{figure}

%\bibliographystyle{utphys}
%\bibliography{references} 

\end{document}